# Influence of multi-line CO laser focusing on broadband sum-frequency generation in ZnGeP$_2$

A.A. IONIN, I.O. KINYAEVSKIY*, YU.M. KLIMACHEV, A.A. KOTKOV, AND L.V. SELEZNEV

*P.N. Lebedev Physical Institute of the Russian Academy of Sciences, Moscow 119991, Russia*
*Corresponding author: kigor@sci.lebedev.ru*

**An influence of multi-line CO laser focusing on spectral characteristics of broadband sum-frequency generation in ZnGeP$_2$ nonlinear crystal was experimentally and numerically studied. Maximal frequency conversion was experimentally observed under tight focusing of the multi-line rather than single-line CO laser. The tight focusing resulted in broadening sum-frequency generation spectrum and increasing total, i.e. integrated over the spectrum, frequency conversion efficiency. These effects were due to increasing phase-matching bandwidth and angular dispersion. Maximal conversion efficiency of the multi-line CO laser was numerically demonstrated to take place at the focal length of 0.4 times that for the single-line one.**

http://dx.doi.org/10.1364/OL.99.099999

A development of mid-IR lasers is very important for gas analysis, laser chemistry, medical and other applications [1]. A carbon monoxide laser (CO laser) is a unique gas laser that can operate in mid-IR on hundreds rotational-vibrational transitions in the wavelength interval from 4.6 [2] to 8.7 μm [3] in a single-line- or multi-line mode simultaneously emitting dozens of spectral lines (see, for instance [4]). Rich spectrum and high power of CO laser make it very attractive for covering mid-IR range by frequency conversion of CO laser radiation in non-linear crystals. For instance, a laser system consisting of CO and CO$_2$ lasers and non-linear crystals operating within wavelength range from 2.5 to 16.6 μm was developed [4]. Such IR laser source would be useful in a variety of researches and applications including gas analysis and the atmosphere sensing [5].

One more attractive point for CO laser research and applications is a noncritical spectral phase-matching in widely-known mid-IR nonlinear crystals such as ZnGeP$_2$, GaSe, AgGaSe$_2$ [6]. Due to noncritical spectral phase-matching, the broadband two-stage frequency conversion of multi-line CO laser radiation was obtained in ZnGeP$_2$ [7, 8] and AgGaSe$_2$ [9] crystals. Similar method of broadband phase-matching frequency conversion was suggested for generation and amplification of ultra-short mid-IR laser pulses [10-12], which has many important applications, such as high harmonic generation [13], for instance.

Phase-matching bandwidth of nonlinear crystal can be even more enhanced by introducing spectral angular dispersion. In this case the different spectral components propagate under different phase matching angles in nonlinear crystal [14-16], i. e. under so-called achromatic phase-matching. A simple way to introduce spectral angular dispersion is laser radiation tight focusing in nonlinear crystal. The achromatic phase-matching was applied for second harmonic generation of both multi-line CO$_2$ laser radiation [14] and femtosecond pulses [15, 16]. In this letter we studied an influence of laser beam focusing on spectral characteristics of broadband sum-frequency generation (SFG) of multi-line CO laser in ZnGeP$_2$, crystal under non-critical spectral phase-matching, which allowed us to control SFG radiation parameters by lens changing. The obtained results may be useful for generation and amplification of ultra-short broadband mid-IR laser pulses as well.

Our experiment was carried out at the Gas Lasers Laboratory of the P.N. Lebedev Physical Institute. The optical scheme of the experiment presented in Fig. 1 was the same as in [8].

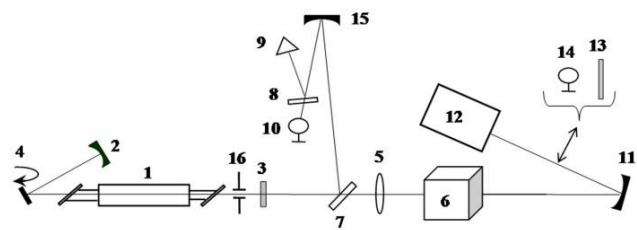

Fig. 1. Optical scheme of the experiment. See text for details.

Low-pressure cryogenic CO laser tube *1* was pumped by a DC discharge. Laser cavity of ~3 m length was formed by spherical mirror

2 (radius of curvature of 9 m) and output mirror *3*. The laser operated in Q-switch mode by rotating mirror *4*. The FWHM pulse duration of the CO laser equaled 0.5 µs at Q-switch frequency of 100 Hz. Peak power of the laser radiation reached 4 kW at average power of 0.2 W. A spectrum of the CO laser consisted of about 115 spectral lines in the wavelength range from 4.9 µm to 6.5 µm (Fig. 2). The CO laser operated on the fundamental transverse mode due to diaphragm *16* 10 mm in diameter installed nearby the output mirror.

Multi-line CO laser radiation was focused by CaF$_2$ lens *5* with the focal length (*f*) of 20 cm or 12 cm onto ZnGeP$_2$ crystal *6* of 15 mm length. The crystal was installed at 47.2° phase-matching angle which corresponded to maximal SFG efficiency. To measure parameters of the CO laser radiation, small part of the laser beam (~5%) was split off by flat CaF$_2$ plates *7* and *8* and directed onto photo-detector PEM-L-3 *9* and laser power meter Ophir-3A *10* with additional spherical mirror *15*. Radiation coming out of the crystal was collimated by spherical mirror *11* and directed to IR spectrometer *12* (IKS-31, LOMO PLC) equipped by a low-noise cryogenic Ge:Au photo-detector or THORLABS PDA20H-EC photo-detector. Measured spectra were displayed by oscilloscope Tektronix TDS5052B. To select and measure power of SFG radiation, fused silica filter *13* and laser power meter *14* Ophir-3A were installed in front of the spectrometer *12*.

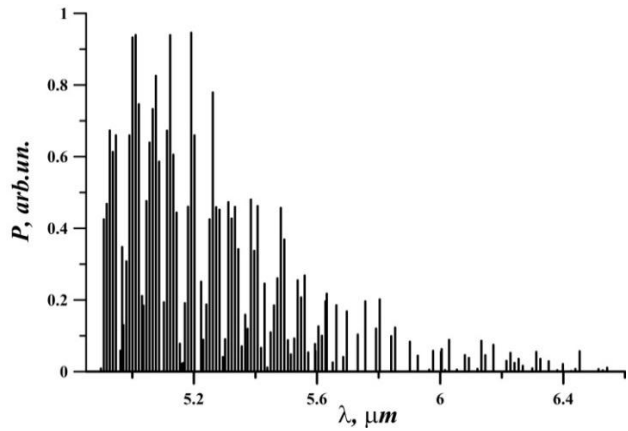

Fig. 2. Spectrum of CO laser.

When the focal length of lens *5* was 12 cm (numerical aperture ~0.04), SFG spectrum consisted of 110 lines in the wavelength range from 2.52 to 2.85 µm (Fig. 3a). External conversion efficiency (the ratio of the SFG radiation power to CO laser power) was 1.8%. When the focal length of lens was 20 cm (numerical aperture 0.025) the SFG spectrum consisted of 50 lines in the wavelength range from 2.55 to 2.67 µm. External conversion efficiency was 1.2%. A comparison of SFG spectra obtained under different focusing shows that increasing numerical aperture by factor ~1.7 broadens SFG spectrum by factor ~2.5 and enhances the number of spectral lines twofold. Peak power of the strongest spectral lines falls down about twofold, but total, i.e. integrated over the whole spectrum, conversion efficiency is 1.5 times higher.

To explain the obtained results, a simplified numerical simulation of the experiment was carried out. Calculated SFG spectra for the focal lengths of 20 cm and 12 cm are presented in Fig. 3b. Under simplified approximation of a plane wave and low conversion efficiency, the SFG radiation power $P_{SF}$ can be calculated by expression [17]

$$P_{SF} = \frac{8\pi^2 d_{eff}^2 L^2 P_{F1} P_{F2}}{\varepsilon_0 c n_{F1} n_{F2} n_{SF} \lambda_{SF}^2 A} sinc^2\left(\frac{|\Delta k| L}{2}\right) \quad (1)$$

where $d_{eff}$ is the effective nonlinear coefficient; $n$ is refractive index; $\lambda_{SF}$ is SFG wavelength; $\varepsilon_0$ is the dielectric constant; $P_{F1}$ and $P_{F2}$ are the pump laser radiation power for two CO laser lines; $\Delta k$ is the wave mismatch, $L$ is the crystal length; $A$ is a cross-section of the laser beam. Dispersion equations and effective nonlinear coefficient were taken from [17]. Placing a lens into optical path results in changing both effective length of crystal and $A$ which affects on conversion efficiency. Also focusing affects on phase-(mis)matching between different CO laser lines through the *sinc* function.

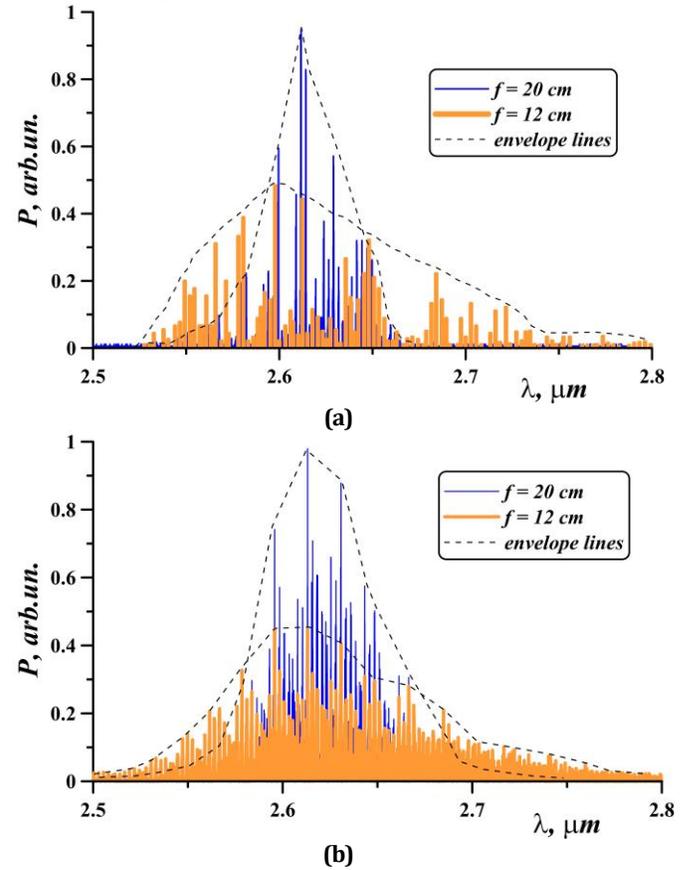

Fig. 3. Measured (a) and calculated (b) SFG spectra for the focal length of 20 cm and 12 cm, dotted lines are envelope lines.

Intensity distribution in the crystal for a tightly focused laser beam should be taken into account. Exact solution of parametric interaction of a Gaussian light beam for single-line radiation was presented in [18, 19]. Our simplified model is close to the theory of second harmonic generation by focused laser beams considered by Bjorkholm in [18]. We considered a Gaussian beam $w(z)$ focused in the middle of ZnGeP$_2$ crystal ($z$=0) of 15 mm length (Fig. 4). The spot radius $w(z)$ of the laser beam is given by

$$w(z) = w_0 \sqrt{1 + \left(\frac{\lambda \cdot z}{n \cdot \pi \cdot w_0^2}\right)^2} \quad (2)$$

where $w_0$ is the waist radius. For simplicity Gaussian beam was calculated for wavelength λ=5.2 µm and refractive index $n$=3.1. Solid and dash lines in Fig. 4 correspond to laser beams with and without nonlinear crystal. The laser beam waist in the crystal was lengthening by factor 3.1 corresponding to refractive index of the ZnGeP$_2$ crystal.

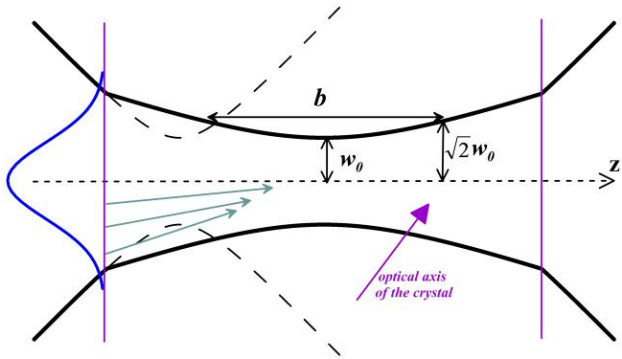

Fig. 4 Gaussian beams with (solid line) and without (dash line) nonlinear crystal.

The calculation of SFG spectra was carried out only for the beam waist because it is most intensive and essential part of interaction. Unlike [18] we did not take into account walk-off effect because of a low $ZnGeP_2$ birefringence ~0.04. By other words, we considered the length of crystal $L$ equaled to the waist length $b$ calculated as:

$$b = 2\frac{n \cdot \pi \cdot w_0^2}{\lambda} \quad (3)$$

The waist edge was defined by expression (2): $w(b/2)$ was $\sqrt{2}$ larger than $w_0$. The cross-section of the laser beam $A$ was calculated as $\pi w_0^2$. Local angular convergence and divergence of the laser beam were calculated as a derivative of the $w(z)$ function in the points of waist edge. To take into account angular spread of wave-vectors, i.e. angular dispersion, the laser beam was divided into small slices in such a way that an angular aperture of the slices was less than the angular phase-matching bandwidth. These slices propagated in the crystal under different angles relative to the optical axes of both the laser beam and nonlinear crystal (see Fig. 4). SFG spectrum for such a slice of the multi-line CO laser beam was calculated by expression (1) taking into account the experimental conditions, as it was made in [8]. The resulted SFG spectrum was calculated as a sum of spectral contributions of these small beam slices.

Calculated SFG spectra for the focal length of 20 cm (which corresponded to equality of the waist and crystal length) and 12 cm (which corresponded to tight focusing) are presented in Fig. 3b. For 12 cm focal length, the waist radius of the CO laser beam equaled 40 μm; the waist length and angular spread were 5.5 mm and ±0.59°, respectively. Calculated SFG spectrum consisted of 939 lines in the wavelength range from 2.52 to 2.76 μm at 0.1 level of maximal power. For 20 cm focal length, the waist radius of the CO laser beam equaled 67 μm; the waist length and angular spread were 15 mm and ±0.35°, respectively. Calculated SFG spectrum consisted of 507 lines in the wavelength range from 2.56 to 2.69 μm at 0.1 level of maximal power.

In our simulation an enhancement of numerical aperture by factor ~1.7 broadened SFG spectrum by factor ~2, enhanced the number of spectral lines by factor ~2, and decreased peak power of the strongest spectral lines also by factor ~ 2, that do correspond to our experimental results. A difference between the number of spectral lines in the experiment and our calculations was connected with limited spectral resolution of the spectrometer and absorption of SFG radiation in the atmosphere. The path length of the laser beam from the crystal to the photo-detector of spectrometer was 5 m. We should have included to the simulation non-collinear SFG between different small beam slices, but even not doing so we got quite a satisfactory agreement between the calculations and the experiment (Fig. 3). This fact demonstrates the applicability of our simple simulation model.

A dependence of frequency conversion efficiency of multi-line and single-line CO laser radiation versus the focal length was also calculated (Fig. 5). The calculated results in Fig. 5 were normalized for visual comparison. For single-line radiation the maximal frequency conversion efficiency was obtained at the focal length of 20 cm which corresponded to equality of the waist and crystal length (15 mm). In our simulation, the making the beam waist length longer and shorter than the crystal length decreases efficiency as proportional to $f$.

However, for the multi-line CO laser we observed quite a different situation. When the focal length decreased from $f = 20$ cm downward, the power of the strongest spectral lines fell down quite sharply. But the total frequency conversion efficiency (integrated over the whole spectrum) was going up and reached maximum at the focal length of 8 cm. It was due to SFG spectrum broadening under tight focusing which compensated power reduction on particular lines. SFG spectrum broadening was caused by two factors: 1) phase-matching bandwidth is much broader at lesser crystal length; 2) an increase of angular dispersion. In the experiment we did get higher efficiency at the lesser focal length.

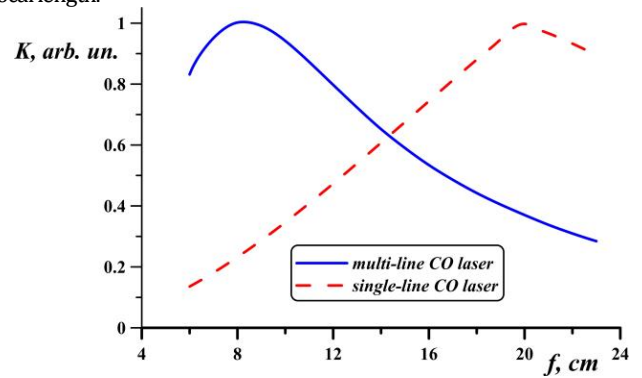

Fig. 5. Frequency conversion efficiency for multi-line (solid line) and single-line CO laser (dash line) radiation in 15-mm length $ZnGeP_2$ crystal versus focal length.

Thus, frequency conversion of multi-line CO laser radiation in $ZnGeP_2$ nonlinear crystal differs significantly from that of single-line one. For multi-line radiation tight focusing appeared to be more efficient. The tight focusing resulted in broadening sum- frequency generation spectrum and increasing total frequency conversion efficiency (integrated over the spectrum). SFG spectrum broadening under tight focusing was caused by two factors: by broader phase-matching bandwidth at lesser effective crystal length and angular dispersion enhancement. Adequate agreement between envelopes of the experimental and calculated spectra demonstrates applicability of our simple simulation model.

**Funding.** Russian Science Foundation (Project No 16-19-10619)